\documentclass[12pt,preprint]{aastex}
\slugcomment{to appear in  ApJ Letters, 10 August 2004}

\begin{document}

\title{Discovery of a candidate inner Oort cloud planetoid}
\author{Michael E. Brown}
\affil{Division of Geological and Planetary Sciences, California Institute of Technology, Pasadena, California 91125}
\email{mbrown@caltech.edu}
\author{Chadwick Trujillo}
\affil{Gemini Observatory}
\email{trujillo@gemini.edu}
\author{David Rabinowitz}
\affil{Yale University}
\email{david.rabinowitz@yale.edu}
\begin{abstract}
We report the discovery of the minor planet 2003 VB12 (popularly named
Sedna), the most
distant object ever seen in the solar system. Pre-discovery images from 
2001, 2002, and 2003 have allowed us to refine the orbit sufficiently
to conclude that 2003 VB12 is on a highly eccentric orbit which 
permanently resides well beyond the Kuiper belt with a semimajor axis of
480$\pm$40 AU and a perihelion of 76$\pm$4AU. Such an orbit is unexpected
in our current understanding of the solar system, but could be the result
of scattering by a yet-to-be-discovered planet, perturbation by an anomalously
close stellar encounter, or formation of the solar system within a cluster of
stars. In all of these cases a significant additional population is likely 
present, and in the two most likely cases
2003 VB12 is best considered a member of the inner Oort cloud, which
then extends to much smaller semimajor axes than previously expected.
Continued discovery and orbital characterization of objects in this inner 
Oort cloud will verify the genesis of this unexpected population.
\end{abstract}

\section{Introduction}
The planetary region of the solar system, defined as the region that 
includes nearly circular low inclination orbits, 
appears to end
at a distance of about 50AU from the sun at the edge of the classical
Kuiper belt
(Allen et al. 2001, Trujillo and Brown 2001). Many high eccentricity bodies from the planetary region -- 
comets and scattered
Kuiper belt objects -- cross this boundary, but all have perihelia 
well within the planetary region.
Far beyond this edge lies the realm of
comets, which are hypothesized to be stored at distances of 
$\sim 10^4$ AU in the Oort cloud.  While many objects 
presumably reside
in this Oort cloud indefinitely, perturbation
by passing stars or
galactic tides occasionally modifies the orbit of a small number of these
Oort cloud objects,
causing them to reenter the inner solar system where they are detected
as dynamically new comets (Oort 1950, Duncan et al. 1987), allowing a dynamical glimpse into 
the distant region from which they came.
Every known and expected
object in the solar system has either a perihelion in
the planetary region, an aphelion in the Oort cloud region, or both.

Since November
2001 we have been systematically surveying the sky in search of distant
slowly moving objects using the Samuel Oschin 48-inch Schmidt
Telescope at Palomar Observatory (Trujillo and Brown 2004) 
and the Palomar-Quest large-area CCD camera
(Rabinowitz et al. 2003).
This survey is designed to cover the majority 
of the sky visible from Palomar over the course of approximately 5 years
and, when finished, it will be the largest survey for distant
moving objects since
that of Tombaugh (1961).
The major goal of the survey is to discover rare large objects in the Kuiper
belt which are missed in the smaller but deeper surveys which find the 
majority of the fainter Kuiper belt objects (i.e, Millis et al. 2001).

In the course of this survey we
detected an object with an R magnitude of 20.7 on 14 November 2003
which moved 4.6 arcseconds over the course of 3 images
separated by a total of 3.1 hours (Figure 1). 
Over such short time periods, the motion of an
object near opposition
in the outer solar system is dominated by the parallax caused by
the Earth's motion, so we can estimate that 
$R \approx 150/\Delta$, where $R$ is the heliocentric distance of the
object in AU and $\Delta$ is the speed in arcseconds per hour. From this 
estimate we can immediately conclude that the detected object is at a distance
of $\sim$100AU, significantly beyond the 50 AU planetary region, and more
distant than any object yet seen in the solar system. The object has been
temporarily designated minor planet 2003 VB12.

Followup observations from the Tenagra IV telescope, the Keck Observatory, 
and the 
1.3-m SMARTS telescope at Cerro Tololo
 between 20 November 2003 and 31 December 2003 
\footnote{see http://cfa-www.harvard.edu/mpec/K04/K04E45.html for a table of
astrometric positions.}
allow us to compute a preliminary orbit for the object  using
both 
the method of Bernstein and Khushalani (2000; hereafter BK2000), 
which is optimized for distant objects in
the solar system, and a full least-squares method which makes no a priori
assumptions about the orbit\footnote{see http://www.projectpluto.com/find\_orb.html}. Both methods suggest a distant eccentric 
orbit with the object currently near perihelion, but derived values for the
semimajor axis and eccentricity are very different, showing the limitations
of fitting an orbit for a slowly moving object with such a small orbital arc.
For such
objects a time baseline of several years is generally required before
an accurate orbit can be determined. 

\section{Pre-discovery images}
For sufficiently bright objects, like the one discovered here,
observations can frequently be found in archival data to extend the time
baseline 
backwards in time.
At each time that a
new position in the past is found a new orbit is computed and earlier 
observations can then be sought.

The object should have been observed on 30 August and
29 September 2003 during drift-scans from the Palomar-QUEST survey 
Synoptic Sky Survey (Mahabal et al. 2003) 
also operating on the
Samuel Oschin telescope at Palomar Observatory. 
From the November and December data we predict
positions for 29 September with an error ellipse of only 1.2 by 0.8
arcseconds (though the two orbital determination methods disagree on precise
orbital parameters, they both predict the same position within an arcsecond). A single
object of the correct magnitude
appears on the Palomar-QUEST images within the error ellipse (Fig 2). A search of
other available archival sources of images of this precise region of the sky, 
including
our own survey data, 
additional Palomar-QUEST data taken on different nights, the Palomar Digitized
Sky Survey images, and the NEAT Skymorph data base\footnote{see 
http://skyview.gsfc.nasa.gov/skymorph}
finds no
object that has ever appeared at this position at any other time.
Below we will refer to such detections which are seen on one date only as
``unique detections.''
Unfortunately, individual images in the Palomar-QUEST survey are not taken
long enough apart for us to determine if this object is moving or is instead
a fixed source which was coincidentally bright only
during the time of observation (a variable star, a supernova, etc.). 
We estimate the probability of an accidental unique detection within
the error ellipse by
examining the 5 by 5 arcminute region surrounding this object to see
if additional unique detections randomly occur. We find no such unique
detections in the surrounding region,
thus the probability of such a unique detection
randomly occurring within the error ellipse appears less than $10^{-4}$.
We conclude that this detection is indeed a pre-discovery image of 2003 VB12.

Including this position 
in our orbit calculation
shrinks the error ellipse for 30 August 2003 -- another 
night of Palomar-QUEST observations -- to less than an arcsecond.
Examination of the 30 August 2003 Palomar-QUEST image and other archival
images of the same location shows a unique detection
at precisely the predicted location. Again, no other unique detection
is found within a 5 by 5 arcminute surrounding box.
We again conclude that
this is our object with very low probability of coincidence.

From a four month baseline the orbital elements are still
uncertain, but positions for the 2002 season 
can be predicted with reasonable 
accuracy.
A search of the Skymorph database of NEAT observations shows that
high-quality images were obtained surrounding the predicted location of our
object from the Samuel Oschin telescope
on the nights of 9 and 29 October 2002. The two orbital prediction methods
described above predict positions separated by 8.5 arcseconds, though the
BK2000 method suggests an error ellipse of semimajor axis only 4.2 arcseconds.
This positional discrepancy is caused by an energy constraint 
in the BK2000 method which
breaks degeneracies in short-arc orbits by preferring lower energy
less eccentric orbits. 
The least-squares method, with no such constraint, finds a more eccentric
orbit and therefore a slightly different position. We estimate an error ellipse
for the least-squares method by a Monte Carlo method in which we add 0.3
arcsecond errors to our observations and recalculate an orbit and predicted
position. 

Figure 2 shows the 29 October 2002 NEAT data with both 
predictions and error ellipses.
A single unique detection of the right magnitude
appears within the full 5 by 5 arcminute field
shown, and this detection is well within the error ellipse of the 
more eccentric least-squares orbital fit. The probability of the single
unique detection randomly 
falling within either error ellipse is $5 \times 10^{-4}$.
Including this detection
in our fit breaks the orbital degeneracy, and now the BK2000 and least-squares 
method find essentially the same orbit and same errors.
With the inclusion of the 29 October point, the error 
for 9 October 2002 shrinks to less than an arcsecond.
Again, the only proper magnitude
unique detection within a 5 by 5 arcminute area
appears at precisely this location and we are confident that we have
detected 2003 VB12. 

Extension of the orbit to 2001 yields additional potential detections
from the NEAT survey on 24 October and 26 September. The 24 October error
ellipse is 2.1 by 0.7 arcseconds, and a unique detection of the correct
magnitude appears within
this small area.
The data quality in 2001 is not as high as the previous data and this
detection is near the limit of the images. Consequently the 5 by 5 arcminute
surrounding area contains 3 additional unique detections of approximately the
same magnitude. Nonetheless, the probability is only $1.5 \times 10^{-3}$ 
of one of these random unique detections falling within our small error ellipse.
The 26 September data contains a unique detection at precisely the right
location, but also 3 other comparable unique detections within 5 arcminutes.
The random probability is less than $10^{-3}$. We conclude that both 2001
images indeed show our object.

Attempting to propagate the orbit to 2000 or earlier results in several
potential detections but the data quality are sufficiently low that
we deem the probability of coincidence too high to consider these.
A special attempt was made to find the object in September
1991 Palomar Digitized
Sky Survey images where the error ellipse is still only 26.7 by 1.1 arcseconds
and while a unique detection can be found within the error ellipse,
we find many potentially spurious unique detections at the same level
and determine the probability for such a random detection to be as
high as $\sim 3$\%, so we discount this candidate early detection as
unreliable.

\section{Orbital solution}
The best fit BK200 orbit for the full set of 
2001-2003 data yields a 
current heliocentric distance ($r$) of $90.32\pm 0.02 $, a semimajor
axis ($a$) of $480\pm 40$AU, an eccentricity $e$ of $0.84\pm 0.01$, 
and an inclination $i$ of 11.927. The object reaches perihelion at a 
distance of $76$ AU on 22 September 2075$\pm$260 days.
The RMS residuals to the best-fit error are 0.4 arcseconds with a maximum of
0.6 arcseconds, consistent with the measurement error of the positions of these
objects. The full least-squares method gives results within these error bars.

The heliocentric distance of 90AU, consistent with the simple estimate
from the night of discovery, is more distant than anything 
previously observed in the solar system. Many known Kuiper belt
objects and comets travel on high eccentricity orbits out to that distance
and beyond, so detection of a distant object is not inconsistent with our
present understanding of the solar system. 
The distant perihelion is, however,
unanticipated. The most distant perihelion distance of any well known
solar system object is 46.6 AU for the Kuiper belt object 1999 CL119.
To verify the robustness of the distant perihelion for 2003 VB12, 
we recomputed 200
orbits while randomly adding 0.8 arcsecond of noise (twice the RMS residuals)
to each of the 
astrometric observations and find that the derived perihelion remains within
the range 73 to 80 AU.

\section{Origin}
The orbit of this object is unlike any other known in the solar system. It
resembles a scattered Kuiper belt object, but with a perihelion much higher 
than can be explained by scattering from any known planet. The only mechanism 
for placing the object into this orbit requires either perturbation by
planets yet to be seen in the
solar system or forces beyond the solar system.
\subsection{Scattering by unseen planet}
Scattered Kuiper belt objects acquire
their high eccentricities through gravitational interaction with
the giant planets. Such
scattering results in a random walk in energy and thus semi-major axis, but
only a small change in perihelion distance. Scattering by Neptune is thought 
to be able to move an object's perihelion only out to distances of $\sim$36 AU.
(Gladman et al. 2002), though more complicated interactions including migration can occasionally
raise perihelia as high as $\sim$50AU (Gomes, 2004), sufficient to explain all
of the known Kuiper belt objects.
Our object could not be scattered
into an orbit with a perihelion distance of 76 AU by any of the major planets.
An alternative, however, is the existence of an undiscovered approximately
earth-massed planet at
a distance of $\sim$70 AU which scattered the object just as Neptune
scatters the Kuiper belt objects. 
Hogg et al. (1991) place dynamical limits the existence of
such a planet and show that a planet at 70 AU of approximately 2 earth masses
should cause detectable modifications of the orbits of the giant planets,
but no dynamical constraints exist on smaller objects.
Nonetheless our current survey has covered at least
80\% of the area within 5 degrees of the ecliptic -- where such a planet 
would be most expected -- with no planetary detections (Brown and Truillo 2004).
We therefore deem the existence of such a scattering planet unlikely, but
we are unable to rule the possibility out completely.

Nonetheless, if such a planet does indeed exist -- or did exist at one time -- 
its signature will be unmistakable in the orbital parameters of
all additional new objects detected in this
region. All should have modest inclinations and perihelion similar to the 76AU
perihelion found here.

\subsection{Single stellar encounter}
This unusual orbit resembles in many ways
one expected for a comet in the Oort cloud.
Oort cloud comets are thought to originate
in the regular solar system where they suffered encounters with
giant planets which scatter them in to highly elliptical orbits. 
When these eccentric
orbits take the comets sufficiently far from the sun, random
gravitational perturbations from passing stars and from galactic tides 
modify the orbit, allowing the perihelion distance to wander and
potentially become decoupled from the regular planetary system. Calculations
including the current expected flux of stellar encounters and galactic tides
show that a comet must reach a semimajor axis of $\sim 10^4$ AU before these
external forces become important (Oort 1950, Fernandez 1997). Once comets obtain such a large semimajor
axis the orbits become essentially thermalized, with mean eccentricities
of 2/3 and isotropic inclinations. Continued perturbations can move the
perihelion back into the planetary region where the object becomes a new
visible comet with a semimajor axis still $\sim 10^4$ AU. 

The major inconsistency between this picture of the formation of the Oort cloud
and the orbit of our newly discovered object is the 
relatively small semimajor axis of the new object
compared to the distance at which forces outside of the solar system should
allow significant perihelion modification. Calculations show that a body
with a semimajor axis of 480 AU and a perihelion in the planetary region
should have had its perihelion modified
by $\la 0.3$\% over its lifetime due to external forces (Fernandez 1997). 
Perihelion modification of such a
tightly bound orbit requires a stellar encounter much closer than expected
in the solar system's current galactic environment.

Only a small range of encounter geometries
are capable of perturbing a scattered Kuiper belt-like orbit to this more Oort
cloud-like orbit. 
As an example, simple orbital integrations show that
an encounter of a solar mass star moving at 30 km s$^{-1}$ perpendicular
to the ecliptic at a distance of 500 AU will perturb an orbit with a 
perihelion of $\sim$30 AU and semimajor axis of $\sim$480 AU to one with a 
perihelion of 76 AU, like that seen. The need for a special geometry is not
surprising, as any single stellar encounter would have a geometry that is
unique. More difficult to explain, however, is that fact that in the present
stellar environment, the probability of even one encounter
the solar system is only about 20\% (Fernandez 1997). 
If the population of objects on large scattered orbits were
in steady state the rarity of such an encounter would matter less, as the 
encounter could occur any time in the past 4.5 billion years. In reality,
however, the number of highly elliptical orbits capable of being perturbed into
the inner Oort cloud must have been significantly higher very early in the
history of the solar system when the outer solar system was being cleared
of icy planetesimals and the Oort cloud was being populated. The
probability of a random
close stellar encounter so early is improbable.

Nonetheless, if such a stellar encounter did indeed occur, its signature 
will be unmistakable in the orbital parameters of all subsequent 
objects found in this region. If all of the objects found in this inner
Oort cloud region are consistent with the same unique stellar encounter geometry
it will be clear that we are seeing the fossilized signature of this 
encounter.

\subsection{Formation in a stellar cluster}

Close encounters with stars would have been more frequent early in the history
of the solar system if the sun had formed inside a stellar cluster.
In addition, these encounters would have been at much slower speeds, leading
to larger dynamical effects.
In numerical simulations, 
Fernandez and Brunini (2000) found that 
early multiple slow moderately close encounters
are capable of perturbing objects into orbits such as the one
here. The process is identical to that hypothesized for the creation of the more
distant Oort cloud, but in a denser environment the comets do not need to have
as large of a semimajor axis before they are perturbed by the stronger
external forces.
Fernandez and Brunini predict a population of
objects with semimajor axes between $\sim 10^2$ and $\sim 10^3$ AU, perihelia 
between $\sim 50$ and $\sim 10^3$ AU, large eccentricities (mean $\sim$0.8), 
and a large inclination
distribution (a full-width-half-maximum of $\sim$ 90 degrees)
in this inner region of the Oort cloud formed in 
an early dense stellar environment.

The inclination of 2003 VB12 appears unusually small 
compared to the large expected inclination
distribution of such an inner Oort cloud population. However an observational
bias exists for detecting objects with inclinations similar to the ecliptic
latitude of the observation. In our observations, 2003 VB12 was discovered
at an ecliptic latitude of 11.9 degrees and has a measured inclination of
11.9 degrees. The probability that an object found at 12 degrees latitude
has an inclination less than 13 degrees if the object is 
drawn from a widely distributed
population like that predicted by Fernandez and Brunini is $\sim$10\%.
 A third of
all objects at 12 degrees will have inclination smaller than 20 degrees.
We thus do not find the small inclination of 2003 VB12 to be inconsistent 
with the distribution expected in this inner Oort cloud scenario.

We currently regard this scenario as the most 
likely for the creation of the unusual orbit of our newly
discovered object. Formation of the solar system in a stellar cluster
is a reasonable expectation (Clarke et al. 2000) for which potential evidence exists from
other contexts (Goswami \& Vanhala 2000).
If indeed this scenario is correct, the orbits of any newly discovered objects
in this region will unmistakably reflect this early history. The new discoveries
will be widely spread in inclination and perihelion and will not be consistent
with any special single stellar encounter geometry. As seen in the simulations
of Fernandez and Bruini, the precise distribution of orbits in this inner
Oort cloud will be indicative of the size of this initial cluster.

It is possible that a second such
object is known already (or perhaps more). 
The scattered Kuiper belt object 2000 CR105 has a
perihelion distance of 44 AU and a semimajor axis of 227 AU. Its present
orbital configuration can be fully explained by a complex path involving
migration of Neptune, scattering, and resonances (Gomes 2004), 
so its existence
does not require any external forces. However,  
the cluster-formation scenario naturally leads to orbits such as that of 
2000 CR105. The relatively small perihelion change of 2000 CR105 in
this scenario is the consistent with the relatively modest semimajor axis
of the object. Unfortunately, 2000 CR105 is close enough to the planetary 
region that it has possibly suffered enough interaction to change its orbital
parameters to erase the clear dynamical signatures we seek in this population.

\section{Discussion}
Each of the plausible scenarios for the origin of the distant object predicts
a specific dynamical population beyond the Kuiper belt. With only a 
single object, though,
little dynamical evidence exists for preferring any one scenario.
With any new discoveries in this region, however, evidence should quickly mount.

We can make a simple
order of magnitude estimate of the ease of future discovery of objects
in this population. We find a single distant object in our survey while
we have found 
40 Kuiper belt objects discovered to date in the survey.
Assuming the size distribution of the distant 
population is the same as that of the Kuiper belt, 
other surveys should find similar
proportions, assuming they are equally sensitive to slow motions. 
As of 15 March 2004, 831 minor planets have been detected beyond Neptune,
we thus expect to have seen $\sim$20 similar objects from other
surveys. Even with this rough estimate, the lack of previous detection appears
significant, suggesting either than most surveys have not been sensitive
to motions as slow as $\sim$1.5 arcseconds per hour or that there is an
overabundance of comparatively bright objects in the distant population.
In either case, it appears likely that new objects in this population should
be detected reasonably soon.

The most plausible scenario for the origin of our object appears to be
the dynamical effect of the creation of the solar system within a dense
stellar cluster. In this scenario the Oort cloud extends from its expected
location at $\sim$100000 AU all the way in to the location of 2003 VB12.
If this scenario is indeed correct the total mass of the
Oort cloud must be many times higher than previously suspected. The expected
population of large objects like the one discovered here is large. 
Our survey could
only have detected this object during $\sim$1\% of its orbit, suggesting a population
of $\sim$100 objects on similar orbits. Moreover, if the population is nearly 
isotropic, $\sim$5 more such objects must be observable 
in the current sky, with a 
total population of 500. 
Assuming a size distribution similar to the Kuiper belt,
the total mass of this population is $\sim$5 earth masses. 
The unseen population with 
ever more distant perihelia are likely even more numerous. With only the single
object known in this population, 
extrapolation of a precise mass is not possible,
nonetheless the existence of a nearby massive previously unsuspected 
inner Oort cloud appears likely. Even in the other origins scenarios a
significant new mass must likely be present. At these distances and, in 
particular for isotropic distributions, current dynamical methods are
unable to rule out any reasonable populations (Hogg et al. 1991). 
If the distant 
populations are sufficiently large, however, they may be detectable in
future occulatation surveys.

While the genesis of 2003 VB12 is currently uncertain, continued discovery
and orbital characterization of similar high perihelion objects should
allow a unique and straightforward interpretation of this population.
Each hypothesized formation mechanism leads to the prediction of a different
dynamically distinct population in the outer solar system.
Study of these populations will
lead to a new knowledge of the
earliest history of formation of the solar system.

\acknowledgments
We thank the staff at Palomar Observatory for their dedicated support
of the the robotic operation of the Samuel Oschin telescope and the 
Palomar-QUEST 
camera. We are grateful to D. Stern, A. Dey, S. Dawson,
and H. Spinrad for obtaining critical followup
observations at Keck Observatory and M. Schwartz of Tenagra Observatory for
making his robotic telescope available for followup. We commend Project Pluto
for making their fine orb\_fit software freely usable by all. Re'em Sari has been
inspirational in discussing scenarios of origin.
This research is supported 
by a Presidential Early Career Award from NASA Planetary Astronomy.

\clearpage

\begin{figure}
\plotone{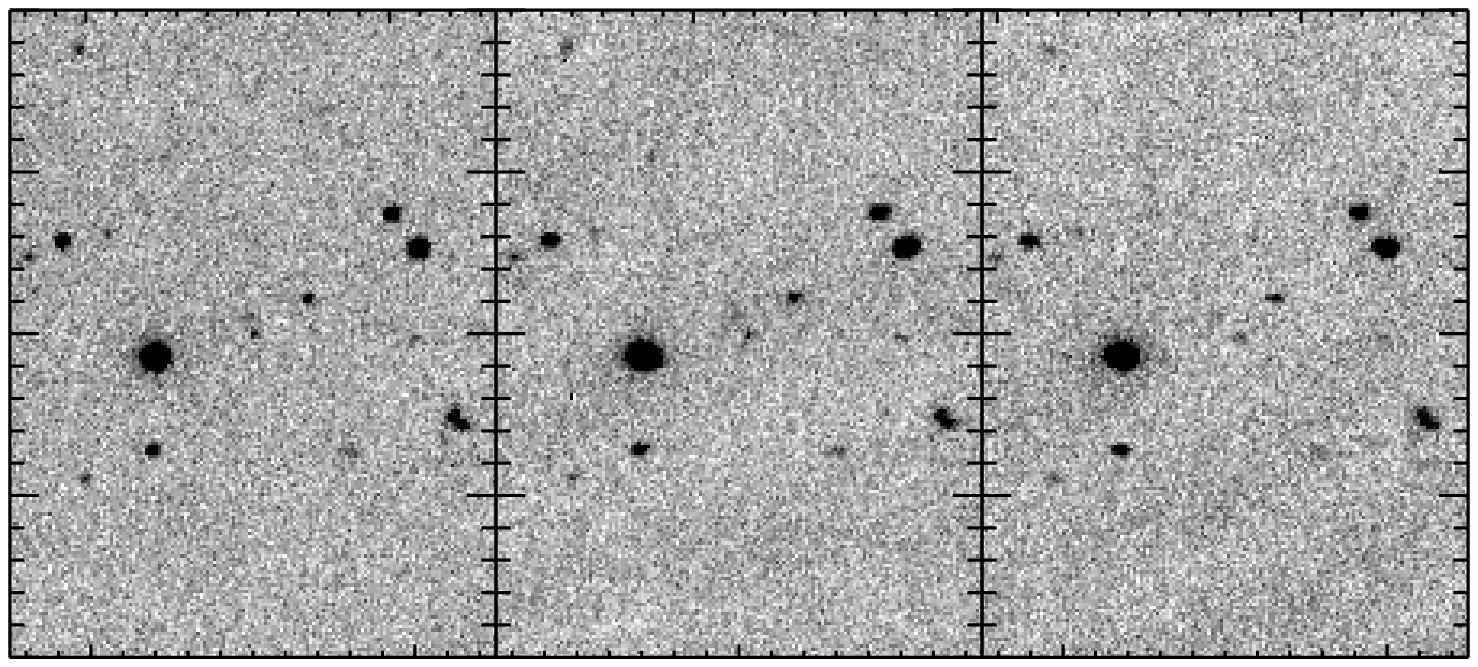}
\caption{Discovery images of 2003 VB12 from the Palomar Samuel Oschin
Telescope and the Palomar-QUEST camera. The pixel scale is 0.9 arcsecond
per pixel with north up and east left. The 150 second exposures
were obtained 14 November
2003 at 6:32, 8:03, and 9:38 (UT), respectively. The object moves 4.6 arcseconds
over 3.1 hours.}
\end{figure}

\begin{figure}
\plotone{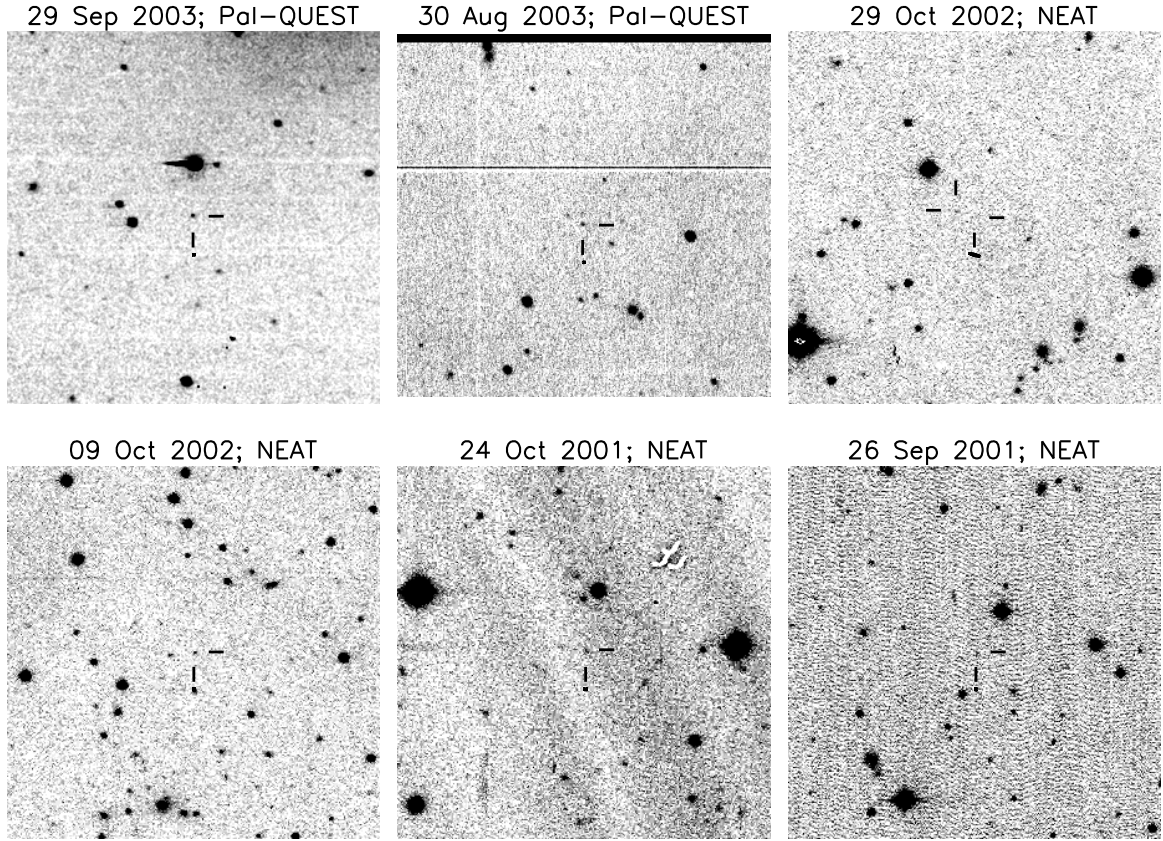}
\caption{Pre-discovery images of 2003 VB12. Each image shows a 5 by 5
arcminute field centered on the predicted position of 2003 VB12. The cross
hairs mark the expected position, while the very small ellipse below the cross
hairs show the size of the error ellipse. In all cases the object
is well within the error ellipse and no similar object appears at the same
position in any other data searched.}
\end{figure}

\end{document}